\documentclass{aa}
\usepackage{graphicx}
\usepackage{natbib}
\usepackage{latexsym}

\begin{document}

\title{Diffuse X-ray emission from the planetary nebula NGC\,7009}

\author{Mart\'{\i}n A.\ Guerrero 
        \and Robert A. Gruendl 
        \and You-Hua Chu}

\institute{Astronomy Department, University of Illinois, 
           1002 W. Green Street, Urbana, IL 61801, USA}

\thanks{
Based on observations obtained with {\sl XMM-Newton}, an ESA science 
mission with instruments and contributions directly funded by ESA 
Member States and NASA. }

\offprints{Mart\'{\i}n A.\ Guerrero, \email{mar@astro.uiuc.edu}}

\date{Received / Accepted }

\abstract{

{\sl XMM-Newton} EPIC observations of the planetary nebula (PN) 
NGC\,7009, the Saturn Nebula, have detected extended X-ray emission 
in its central cavity. 
The diffuse X-ray emission must originate in the shocked fast 
stellar wind.
The spectra show that the temperature of the hot gas is 
1.8~10$^6$~K.
The rms density derived from the volume emission measure is a few tens 
H-atom~cm$^{-3}$.
The hot gas does not appear over-pressurized with respect to the 
nebular shell.
NGC\,7009 represents an evolutionary stage at which the 
influence of the hot gas in the PN interior starts to decline 
due to the diminishing strength of the fast stellar wind and the 
expansion of the central cavity.  

\keywords{
(ISM:) planetary nebulae: general -- 
(ISM:) planetary nebulae: individual: NGC\,7009 -- 
stars: winds, outflows -- 
X-rays: ISM
} 
}

\authorrunning{Guerrero et al.}

\maketitle

\section{Introduction}

It has been commonly accepted that the shell structure of a 
planetary nebula (PN) is formed by the current fast stellar wind 
sweeping up the previous asymptotic giant branch (AGB) wind 
\citep[e.g.][]{Kwok83}. 
In this interacting-stellar-winds model, a PN will comprise a central 
cavity filled with shocked fast wind at temperatures of 10$^6-10^8$ K, 
a dense shell of swept-up AGB wind at 10$^4$ K, and an outer envelope 
of unperturbed expanding AGB wind.
The shocked fast wind has a high temperature but does not emit
appreciably in X-rays because its density is low.
At the interface between the shocked fast wind and the dense nebular 
shell, heat conduction lowers the temperature and mass evaporation 
raises the density of the shocked fast wind, producing optimal 
conditions for X-ray emission \citep{Weav77}.  Thus, soft X-ray 
emission from the shocked fast wind in a PN interior is expected to 
peak near the inner wall of the nebular shell.

The $ROSAT$ X-ray satellite had been used to search
for diffuse X-ray emission from PNe with limited success 
\citep{GCG00}.  
It is only recently that the {\it Chandra X-ray Observatory} has 
unambiguously detected diffuse X-ray emission in the central 
cavities of three young PNe: BD\,+30$^\circ$3639 \citep{Kast00}, 
NGC\,6543 \citep{CGGWK01}, and NGC\,7027 \citep{Kast01}.  
In the case of NGC\,6543 where the X-ray emission is well-resolved, 
a clear limb-brightening is observed.
The plasma temperatures implied by their X-ray spectra are a few 
$10^6$~K.  
{\it Chandra} observations of the evolved PN NGC\,7293 (the Helix 
Nebula), on the other hand, does not show any diffuse X-ray emission.  
These results suggest that the hot gas content of a PN diminishes as 
the PN ages.  
To confirm this hypothesis, PNe at intermediate evolutionary stages 
need to be observed.

We have examined all pointed and serendipitous observations of
PNe in the $ROSAT$ archive and found that NGC\,7009, the Saturn 
Nebula, hosts an X-ray source with spectral characteristics of 
plasma emission.  
NGC\,7009 is more evolved than the three PNe with diffuse X-ray 
emission because of its larger nebular size and dynamical age.  
Consequently, NGC\,7009 is an excellent candidate not only for diffuse 
X-ray emission, but also as a PN at intermediate evolutionary stages.  
We have obtained X-ray observations of NGC\,7009 with the
{\it XMM-Newton Observatory}, and found that the X-ray emission 
is indeed extended.  
This paper reports our analysis of the diffuse X-ray emission from 
NGC\,7009.  

\section{Observations}

NGC\,7009 was observed with the {\it XMM-Newton Observatory} 
in Revolution 255 on 2001 April 30 using the EPIC/MOS1, EPIC/MOS2, 
and EPIC/pn CCD cameras.  
The two EPIC/MOS cameras were operated in the Full-Frame Mode for a 
total exposure time of 39.4 ks, while the EPIC/pn camera was operated 
in the Extended Full-Frame Mode for a total exposure time of 31.6 ks.  
The EPIC/pn observations started about 4 ks later than the EPIC/MOS 
observations because of a larger overhead required by the 
EPIC/pn, but both observations concluded at the same time. 
For all observations, the Medium filter was used.  
At energies below 1 keV, the point spread function (PSF) of the EPIC
instruments has a small core with FWHM $\sim 6''$ but extended 
wings with a half energy width of $\sim 15''$.  
For the EPIC/MOS and EPIC/pn CCDs, the pixel sizes are 1\farcs1 and 
4\farcs1, and the energy resolutions are $\sim60$ eV and $\sim100$ eV 
at 1 keV, respectively (Dahlem 1999; Str\"uder et al.\ 2001).  

We received the {\it XMM} pipeline products which included the raw 
and processed data of all instruments.  
These data were further processed using the {\it XMM-Newton} Science 
Analysis Software (SAS version 5.2.0) and the calibration files from 
the Calibration Access Layer available on 2002 January 24.  

The event files were further screened to eliminate bad events, such as 
those due to charged particles, and periods of high background.  
For the EPIC/MOS observations, only events with CCD patterns 0--12 
(similar to $ASCA$ grades 0--4) were selected;  
for the EPIC/pn observation, only events with CCD pattern 0 (single 
pixel events) were selected.
To assess the background rate, we binned the counts over 25 s time 
intervals for each instrument in the 10--12 keV energy range, where 
the counts were dominated by the background. 
The time intervals with high background, i.e.\ count rates $\ge 0.4$ 
cnts~s$^{-1}$  for the EPIC/MOS or $\ge 1.5$ cnts~s$^{-1}$ for the 
EPIC/pn, were discarded.
The resulting exposure times are 35.0 ks, 35.4 ks, and 26.2 ks for 
the EPIC/MOS1, EPIC/MOS2, and EPIC/pn observations, respectively.
A total of $480\pm25$ cnts, $470\pm25$ cnts, and $1,610\pm45$ cnts 
from NGC\,7009 were detected in the EPIC/MOS1, EPIC/MOS2, and EPIC/pn 
observations, respectively.

\section{Spatial Analysis}

We have analyzed the spatial distribution of the X-ray emission 
from NGC\,7009 and compared it to the optical nebula.
As the data analysis is fairly complex, we present images at
various stages of the processing in Figure 1.
First, the event files from the three EPIC instruments are merged 
together to construct a raw EPIC image with a pixel size of 1\farcs5 
in the 0.15--1.25 keV energy band (Fig.~1a).  
As shown in \S 4, this energy band includes all photons detected
from NGC\,7009.
The raw EPIC image is then adaptively smoothed using Gaussian 
profiles with FWHM ranging from 5$''$ to 17$''$.
A grey-scale presentation of this adaptively smoothed EPIC image 
is shown in Fig.~1b.
The contours from this smoothed EPIC image are overplotted on an 
H$\alpha$ image of NGC\,7009 taken with the {\it Hubble Space 
Telescope (HST)} Wide Field Planetary Camera 2 (WFPC2) and presented
in Fig.~1c.
Finally, to improve the spatial resolution, this smoothed EPIC image 
is deconvolved using the Lucy-Richardson algorithm and a theoretical
EPIC PSF provided in the {\it XMM-Newton} Calibration Access Layer.
The contours extracted from the PSF-deconvolved EPIC image are 
overplotted on the {\it HST} WFPC2 H$\alpha$ image in Fig.~1d.

The X-ray emission from NGC\,7009 is clearly extended.  
Its spatial distribution in the adaptively smoothed EPIC image 
(Fig.~1b and 1c) can be described by an elliptical Gaussian with 
FWHM $\sim 23'' \times 14''$ and a major axis along the position
angle $\sim70^\circ$.
There is an excellent correspondence between the X-ray emission
and the nebular shell of NGC\,7009.
The {\it HST} WFPC2  H$\alpha$ image of NGC\,7009 shows a bright
$25'' \times 10''$ inner shell surrounded by a $25'' \times 
20''$ envelope, with two ansae extending from the tips of the 
major axes of the inner elliptical shell out to $25''$ from the 
central star.  
The position and orientation of the 50\% contour of the smoothed EPIC 
image matches almost perfectly the inner shell of the nebula 
(Fig.~1c), indicating that the X-ray emission peaks within the central 
cavity.  
The PSF-deconvolved EPIC image further shows that the majority of the 
X-ray emission originates from the central cavity.  
As the EPIC PSF has extended wings and our PSF-deconvolution is 
limited by the number of counts and the validity of the theoretical 
PSF adopted, it is most likely that all diffuse X-ray emission from 
NGC\,7009 is confined within the inner shell, similar to what is 
observed in BD\,+30$^\circ$3639 and NGC\,6543.  
Furthermore, NGC\,7009 has a sharp-edged inner shell similar to that of 
NGC\,6543, which implies an expansion driven by the pressure of the hot 
gas in the central cavity \citep{CGGWK01}.  
Therefore, we suggest that the hot, X-ray-emitting gas in NGC\,7009 is 
also confined within its central cavity, but the detailed distribution 
of the hot gas cannot be determined from the currently available 
X-ray observations.  
In particular, it is not possible to address whether X-ray emission from 
its central star is present or not.   

\section{Spectral Analysis}

Spectra of NGC\,7009 have been extracted from the event files
of the three {\sl XMM-Newton} EPIC cameras separately because 
they have different instrumental response functions.
A $50\arcsec \times 40\arcsec$ elliptical source aperture is used 
to include all X-ray emission detected from NGC\,7009.  
An elliptical annular region exterior to and $\sim10$ times as 
large as the source aperture is used to assess the background 
level.  
The background-subtracted spectra of NGC\,7009 are shown in 
Fig.~2-{\it left}.

The EPIC/pn spectrum peaks at $\sim0.55$ keV and drops abruptly to a 
faint plateau between 0.7 and 1.0 keV, but stays essentially flat
from 0.5 keV down to 0.3 keV.
The EPIC/MOS spectra also peak at $\sim0.55$ keV, but have 
smaller numbers of counts because of lower sensitivities.
The detailed differences between the EPIC/pn and EPIC/MOS spectra
reflect the different instrumental responses among these cameras
and the low S/N ratios in the latter spectra.
Neither the EPIC/MOS spectra nor the EPIC/pn spectrum shows 
any detectable emission at energies greater than 1.25 keV.  

The EPIC spectra of NGC\,7009 are clearly inconsistent with a 
power-law or blackbody spectrum.
Thus, the X-ray emission from NGC\,7009 must be thermal plasma
emission and the spectral peak at $\sim$ 0.55 keV corresponds 
to the He-like triplet of O~{\sc vii} at $\sim0.57$ keV.
We can model the observed spectra in order to determine the
physical conditions of the X-ray-emitting plasma and the amount 
of foreground absorption.  
We have adopted the thin plasma emission model of \citet{RS77} 
and the nebular abundances determined from optical and UV spectra
of NGC\,7009.
The abundances of He, C, N, O, Ne, and S relative to hydrogen 
by number have been measured to be 0.13, 2.9~10$^{-4}$, 
2.2~10$^{-4}$, 4.3~10$^{-4}$, 7.2~10$^{-5}$, and 1.9~10$^{-5}$, 
respectively (Kwitter \& Henry 1998).  
Accordingly, we have adopted nebular abundances, relative to the solar 
values \citep{AG89}, of 1.33, 0.8, 1.9, 0.7, 0.6, and 1.2 for He, C, 
N, O, Ne, and S, respectively, and 1.0 for the other elements.
We have assumed solar abundances for the foreground interstellar 
absorption, and adopted absorption cross-sections from \citet{BM92}.

The spectral fits are carried out by folding the Raymond-Smith model 
spectrum through the corresponding EPIC/MOS1, EPIC/MOS2, and EPIC/pn 
response matrices, and fitting the modeled spectra to the observed 
EPIC spectra in the 0.3--1.5 keV energy range using the $\chi^2$ 
statistics.  
The best-fit model, overplotted on the EPIC/pn and EPIC/MOS spectra in 
Fig.~2-{\it left}, has a plasma temperature of $T$ = 
1.8~10$^6$ K (or $kT$ = 0.152 keV), an absorption column density 
of $N_{\rm H}$ = 8.1~10$^{20}$~cm$^{-2}$, and a volume emission 
measure of 1.9~10$^{54} d^2$ cm$^{-3}$ where $d$ 
is the distance to NGC\,7009 in kpc\footnote{
The distance estimates to NGC\,7009 range from 0.5 to 2.1 kpc.  
In the following we will adopt a distance of 1.2 kpc.}.

The goodness of the spectral fits is illustrated by the plot
of reduced $\chi^2$ of the fits as a function of $kT$ and 
$N_{\rm H}$ shown in Fig.~2-{\it right}.  
The 99\% confidence contour spans 
$T$ = 1.6--2.0~10$^6$ K (or $kT$ = 0.14--0.17 keV), 
$N_{\rm H}$ = 2--17~10$^{20}$~cm$^{-2}$, and volume emission
measure 1.5--2.4~10$^{54} d^2$ cm$^{-3}$.  
Assuming that the H\,{\sc I} column density is similar to the 
absorption column density and adopting a typical gas-to-dust ratio 
(Bohlin, Savage, \& Drake 1978), the corresponding visual extinction 
is $A_V$ = $0.44^{+0.5}_{-0.3}$ mag, and the logarithmic extinction 
at the H$\beta$ line is $c_{\rm H\beta}$ = $0.15^{+0.22}_{-0.15}$.  
This extinction is consistent with other estimates of extinction 
($c_{\rm H\beta} = 0.24\pm0.06$, Lame \& Pogge 1996; 
$c_{\rm H\beta} = 0.10^{+0.06}_{-0.08}$, Rubin et al.\ 2001). 
The observed (absorbed) X-ray flux of NGC\,7009 in the 0.3--1.5 keV 
energy range is 7.2~10$^{-14}$ ergs~cm$^{-2}$~s$^{-1}$, and the
unabsorbed (intrinsic) X-ray flux is 1.7~10$^{-13}$ 
ergs~cm$^{-2}$~s$^{-1}$. 
The total X-ray luminosity in the 0.3--1.5 keV energy range is 
$L_{\rm X} =$ 2.0~10$^{31} d^2$ ergs~s$^{-1}$ = 3~10$^{31}$ 
ergs~s$^{-1}$ for a distance of 1.2 kpc.

\section{Discussion} 

\noindent 
{\sl XMM-Newton} EPIC observations of NGC\,7009 have shown that
the diffuse X-ray emission is confined within the central cavity.  
The temperature of the X-ray-emitting gas, 1.8~10$^6$~K, is below 
that expected for the shocked fast wind from its central star which 
has a terminal velocity $\sim 2,800$ km~s$^{-1}$ \citep{CP89,IS01}.  
Similar situation is observed in the other PNe with diffuse X-ray 
emission:  the plasma temperatures implied by the X-ray spectra of 
BD\,+30$^\circ$3639 and NGC\,6543 are several times lower than the 
expected post-shock temperature for their fast stellar wind 
\citep{Kast00, CGGWK01}.  
This temperature difference is expected if heat conduction is prevalent 
at the interface between the shocked fast wind and the dense nebular 
shell \citep{S94,ZP96}.  

In order to determine the role played by the hot gas in the nebular 
dynamics of NGC\,7009, it is possible to derive the rms density and 
pressure of the X-ray-emitting gas and compare them to those of the 
surrounding cool nebular shell.  
Assuming a prolate ellipsoidal central cavity, the volume occupied 
by the X-ray-emitting gas is $\sim$6.7~10$^{51} (d/1.2)^3 \epsilon$ 
cm$^3$, where $\epsilon$ is the filling factor.  
From this volume and the aforementioned volume emission measure 
we derive an rms density of 20~$(d/1.2)^{-1/2}~\epsilon^{-1/2}$ 
cm$^{-3}$, and a total mass for the X-ray-emitting gas of 
8~10$^{-5}~(d/1.2)^{5/2}~\epsilon^{1/2} M_\odot$. 
This rms density and the plasma temperature of 1.8~10$^6$~K imply that
the thermal pressure of the X-ray-emitting gas is 
$\sim$4~10$^{-9}~(d/1.2)^{-1/2}~\epsilon^{-1/2}$ dynes~cm$^{-2}$.
The nebular shell surrounding the hot gas has a density of 5,000 
cm$^{-3}$, a temperature of 10,500 K (Hyung \& Aller 1995; Rubin 
et al.\ 2001), and thus a thermal pressure of 
$\sim$10$^{-8}$ dynes~cm$^{-2}$.  
The thermal pressure of the nebular shell is comparable or higher
than that of the X-ray-emitting gas unless the filling factor
$\epsilon$ is $\ll$0.2.  

The X-ray luminosity of NGC\,7009 and the density and pressure of the 
hot gas in its interior are the lowest among the four PNe for wich 
diffuse X-ray emission has been detected thus far.  
These differences may be associated with the nebular evolution.  
NGC\,7009 has a large nebular shell, 0.145$\times$0.06 pc in size, 
and dynamical age, $\sim1,700$ yrs for an expansion velocity of 
17~km~s$^{-1}$ \citep{BLA94}.  
Therefore, it appears to be in the most advanced evolutionary stage 
among these four PNe.  
At a more advanced evolutionary stage, NGC\,7293 does not have 
any detectable diffuse X-ray emission (Guerrero et al.\ 2001).  
The {\it Chandra} and {\it XMM-Newton} observations of these five 
PNe suggest that PNe are brightest in X-rays when they are young 
and NGC\,7009 may be at an evolutionary stage when the diffuse 
X-ray emission starts to decline.  
This evolutionary effect is expected as the stellar wind 
diminishes and the central cavity grows.  
More observations of PNe at different evolutionary stages are 
needed to evaluate the duration of the pressure-driven phase in 
PNe and its effects on the nebular dynamics.  
As the evolved PNe are expected to have low X-ray surface 
brightness, the high sensitivity of {\it XMM-Newton} is essential 
in detecting such PNe.

\begin{acknowledgements}
We thank Steve Snowden for advice on the {\it XMM-Newton} data 
reduction, Guillermo Garc\'{\i}a-Segura for enlightening 
discussion on the X-ray emission from PNe, and Rosina Iping, 
George Sonneborn, and Robert H.\ Rubin for useful discussion on 
the physical properties of NGC\,7009.  
We also thank the referee, Joel Kastner, for his constructive 
comments.  
This research was supported by the NASA grant NAG~5-10042.  
\end{acknowledgements}

\begin{figure*}
\centering
  \includegraphics[width=15.5cm]{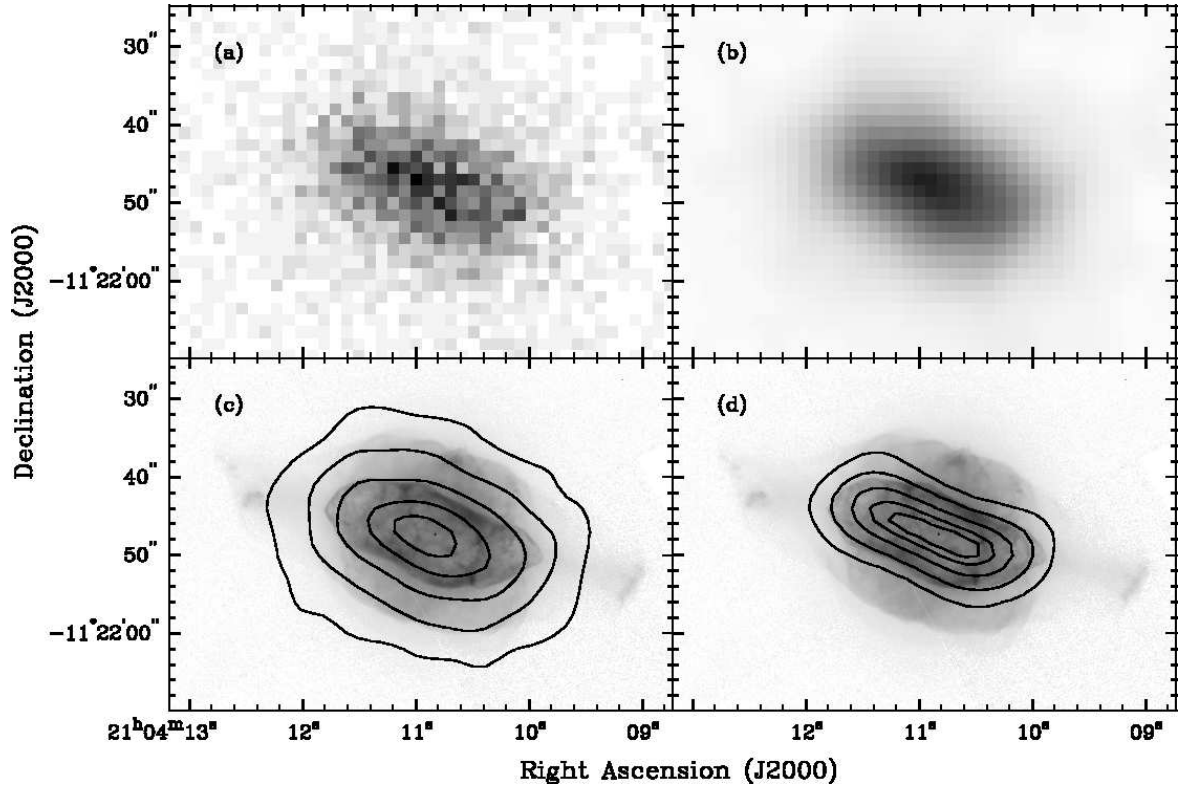}
\caption{
Raw {\it (a)} and smoothed {\it (b)} {\it XMM-Newton} EPIC images of 
NGC\,7009 in the 0.15--1.25 keV energy band.  
X-ray contours from the smoothed EPIC image {\it (c)} and from the 
PSF-deconvolved EPIC image {\it (d)} overplotted on an $HST$ WFPC2 
H$\alpha$ image of NGC\,7009.  
The contour levels are chosen at 10, 25, 50, 75, and 90\% of the emission 
peak.  }
\end{figure*}

\begin{figure*}
\centering
  \includegraphics[width=15.5cm]{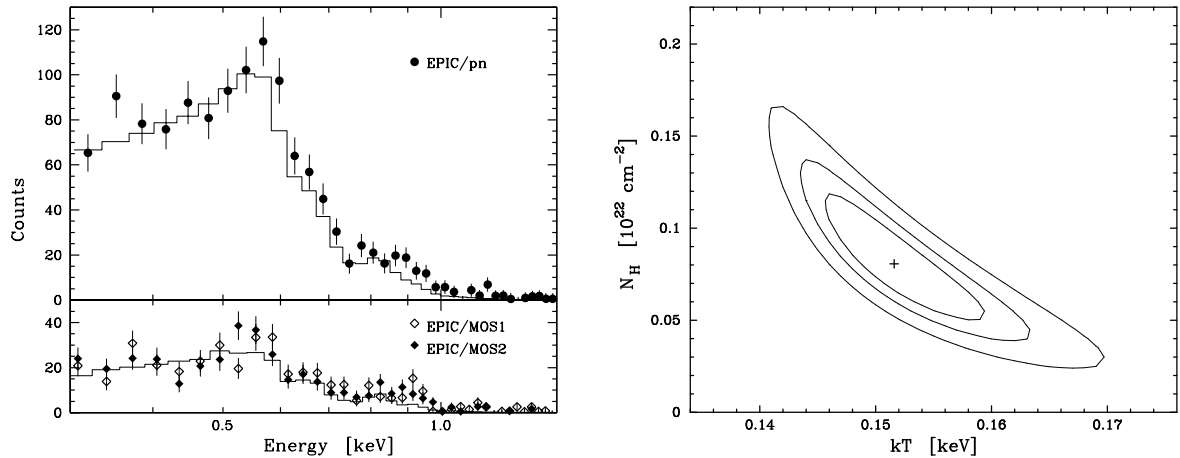}
\caption{
{\it (left)} 
{\it XMM-Newton} EPIC/pn {\it (top)}, and EPIC/MOS1 and EPIC/MOS2 {\it 
(bottom)} background-substracted spectra of NGC\,7009.  
The spectra are overplotted with the best-fit Raymond \& Smith model 
(1977) to the EPIC/pn and the combined EPIC/MOS spectra.  
Both spectra and models are plotted in 30 eV wide energy bins to 
preserve the EPIC spectral resolution ($\sim$ 60--100 eV).  
{\it (right)} 
$\chi^2$ grid plot of the spectral fit of the EPIC/pn and the combined 
EPIC/MOS spectra.  
The contours represent 68\%, 90\%, and 99\% confidence levels, and the 
`+' sign corresponds to the parameters of the best-fit model.
}
\end{figure*}

\end{document}